\documentclass[twocolumn,english,aps,prl, nobalancelastpage,showpacs]{revtex4}
\usepackage{graphicx}
\usepackage{babel}
 

\begin{document}

\title{Alloy effects in ${\rm Ga}_{1-x}{\rm In}_{x}{\rm N}/{\rm GaN}$ heterostructures}

\author{Duc-Phuong~Nguyen}
\email{Duc-Phuong.Nguyen@lpa.ens.fr}

\author{N.~Regnault}

\author{R.~Ferreira}

\author{G.~Bastard}

\affiliation{Laboratoire Pierre Aigrain ENS, 24 rue Lhomond,
75005 Paris, France}

\begin{abstract}
We show that the large band offsets between GaN and InN and the heavy  carrier effective masses preclude the use of the Virtual Crystal Approximation to describe the electronic structure of ${\rm Ga}_{1-x}{\rm In}_{x}{\rm N}/{\rm GaN}$ heterostructures while this approximation works very well for the ${\rm Ga}_{1-x}{\rm In}_{x}{\rm As}/{\rm GaAs}$ heterostructures.
\end{abstract}

\pacs{73.22., 78.40.Fy}

\maketitle

    The use of semiconductor alloys proves to be necessary to adjust the electronic parameters to the design of specific devices. ${\rm Ga}_{1-x}{\rm In}_{x}{\rm N}$ alloys inserted between GaN \cite{Adelmann} would be ideally suited to cover the entire near infrared/ultra-violet spectrum \cite{Nakamura1}. Despite the lack for translation invariance, one may very often describe the electronic behavior of an alloy by means of a band structure, the so-called Virtual Crystal Approximation \cite{Harrison} (VCA).  Close to it, the Coherent Potential Approximation allows to introduce part of the disorder effects in the VCA scheme, which leads to a damping of the VCA Bloch states.  An example of alloys which is well described by the VCA/CPA is ${\rm Ga}_{1-x}{\rm In}_{x}{\rm As}$. The VCA has also been used to describe the electronic properties of ${\rm Ga(In)N}$ heterostructures \cite{Hangleiter}

In this letter, we shall show that the ${\rm Ga}_{1-x}{\rm In}_{x}{\rm N}$ system behaves in a radically different way.  A numerical computation of the electronic states of ${\rm Ga}_{1-x}{\rm In}_{x}{\rm N}/{\rm GaN}$ quantum wells and quantum dots will show very large differences from the VCA predictions. Instead, we shall show that ${\rm Ga}_{1-x}{\rm In}_{x}{\rm As}/{\rm GaAs}$ electronic states are very close to the VCA predictions.  The difference between the two alloy families arises from the much larger band offsets between GaN and InN than those between GaAs and InAs and from the heavier effective masses in the nitride system than in the arsenide system. We note that the large potential offset between ${\rm In}$ and ${\rm Ga}$ has led Kent and Zunger \cite{Kent} to predict that nitride based alloys cannot be described by models which neglect fluctuations.

\begin{figure}[!htbp]
\includegraphics[width=0.42\columnwidth, keepaspectratio]{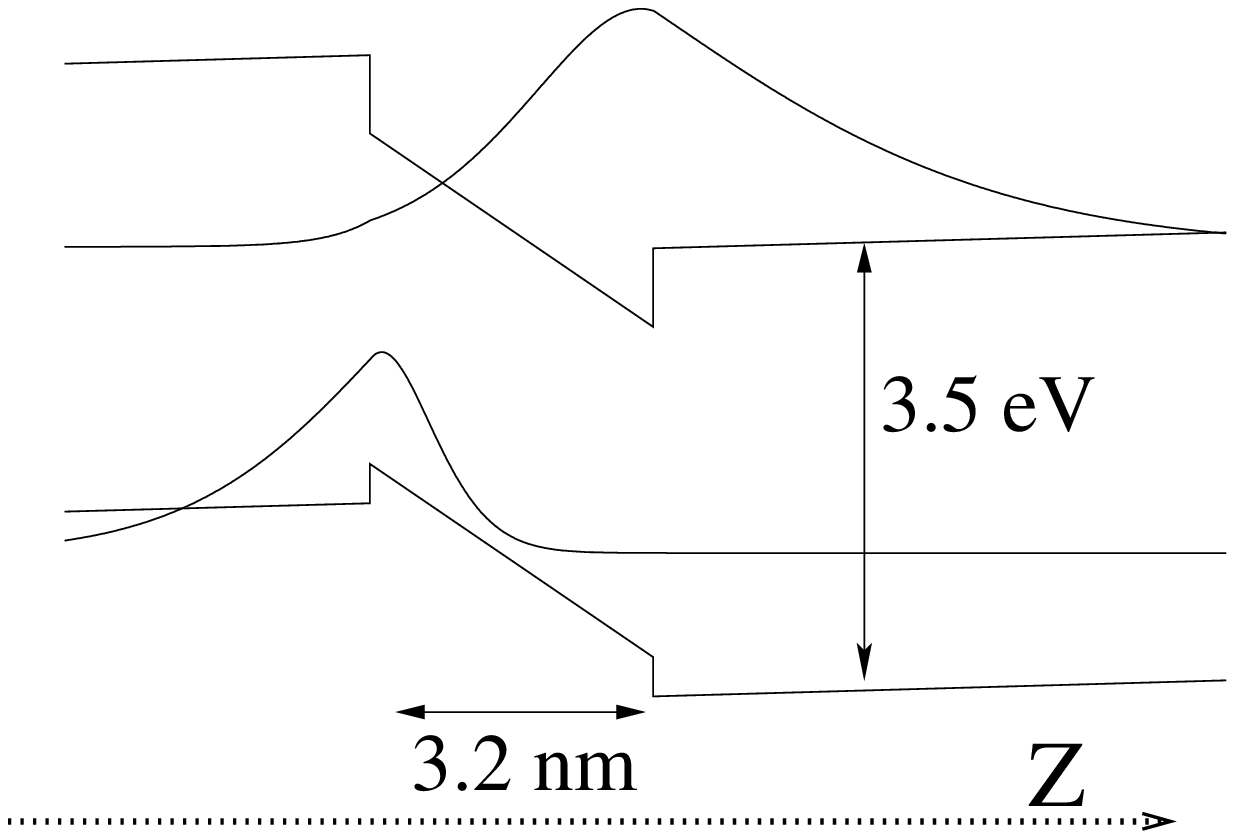}
\vspace{0pt}
\hspace{0.04\columnwidth}
\includegraphics[width=0.42\columnwidth, height=0.35\columnwidth]{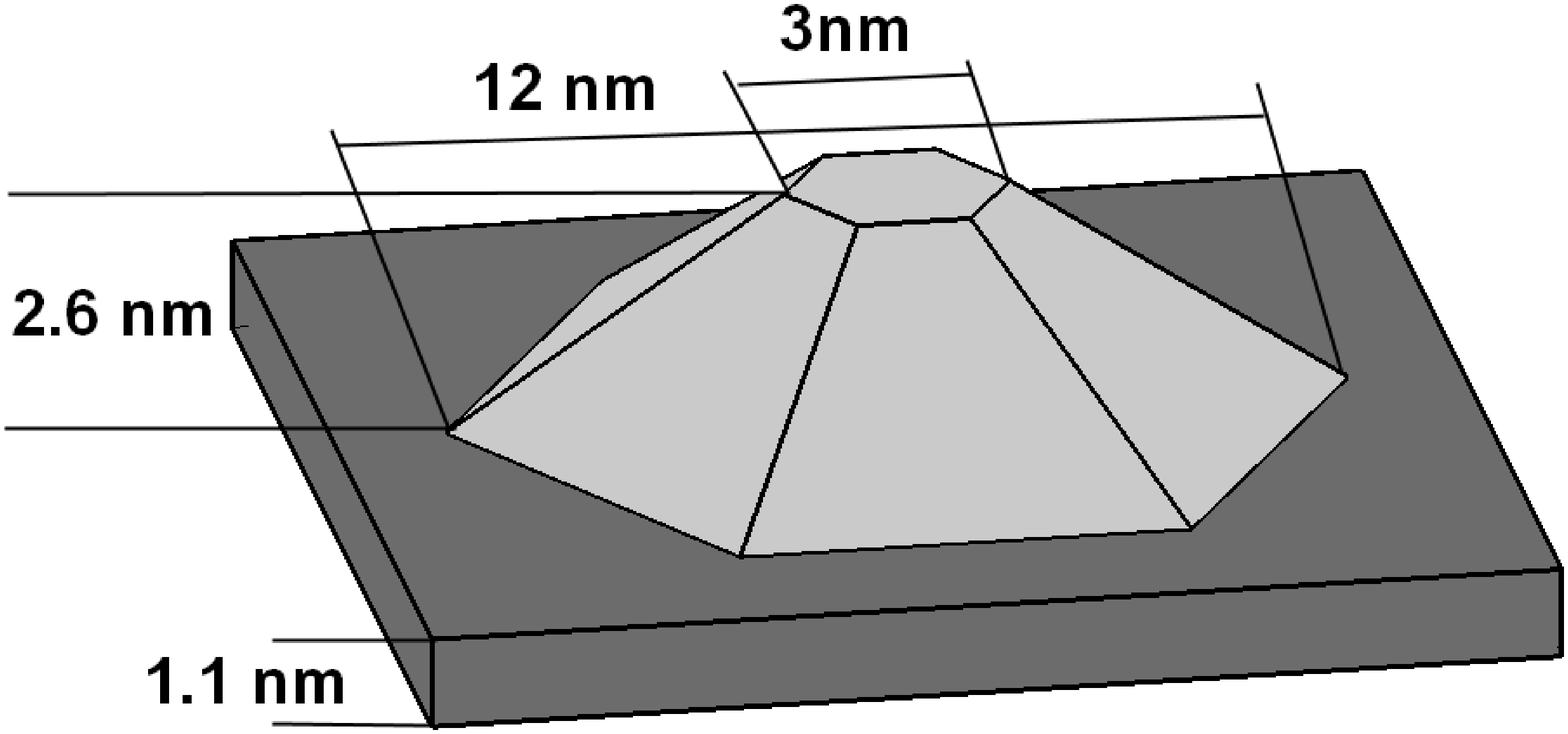}
\caption{Left: VCA band diagram and wave functions in quantum well in the growth direction. Right: geometry of a quantum dot with its wetting layer}
\label{WellDot}
\end{figure}

In the following, we use one band effective mass Hamiltonians to describe the electron and hole kinematics.  The effective mass is slightly anisotropic in the conduction band ($m_{e z} = 0.184\;m_0,\; m_{e //} = 0.166\;m_0$) and anisotropic in the valence band ($m_{h z} = 1.1\;m_0,\; m_{h //} =  0.5045\;m_0$) \cite{Morel1, Morel2}.  The conduction $\Delta E_c$ (valence $\Delta E_v$) band offsets between GaN and InN is taken equal to 1.8~eV  (0.9 eV) while those between GaAs and InAs are taken equal to 0.41 eV and 0.29 eV respectively \cite{Stier}. Thus, there is a much larger energy fluctuation in the nitrides than in the arsenides when in a given unit cell a Ga atom replaces an In atom.  The two heterostructures we shall be dealing with are a 3.2 nm thick ${\rm Ga}_{0.83}{\rm In}_{0.17}{\rm N}/{\rm GaN}$ quantum well and a ${\rm Ga}_{0.83}{\rm In}_{0.17}{\rm N}$ truncated pyramid with hexagonal basis (6 nm side), 2.6 nm height and a basis angle  of  $30^{\circ}$ embedded into a GaN matrix and floating on a  1.1~nm  thick wetting layer (see Fig.(\ref{WellDot})).  A pyramid with the same geometrical parameters made of ${\rm Ga}_{0.5}{\rm In}_{0.5}{\rm As}$ and embedded into GaAs will be considered for comparison (50\% of In instead of 17\% was considered to get bound states to the pyramid).  It is well known that nitride heterostructures contain huge internal fields \cite{Morel1, Morel2, Bernardini}.  This will be modeled by assuming that there exist piecewise constant electric fields which are oriented along the $Z$ direction and have a magnitude of 2.45~MV/cm in ${\rm Ga}_{0.83}{\rm In}_{0.17}{\rm N}$ and - 0.1~MV/cm in GaN \cite{Morel1, Morel2}.

The quantum well problem will be reduced to a two dimensional (2D) calculation by considering an effective in - plane potential which is the average of the actual 3D potential weighted by $\left|\Psi(Z)\right|^2$ where $\Psi\left(Z\right)$ is the envelope function solution of the 1D problem in the presence of the electric field and a square well confining potential with a depth of  0.31 eV and  0.15 eV ($= 0.17\Delta E_c$ or $0.17\Delta E_v$) (see Fig.(\ref{WellDot})). This effective 2D potential is then supplemented by a hard box potential which confines the carrier in the layer plane into a square with dimensions  $L\times L=36\; {\rm nm} \times 36\; {\rm nm}$. A given sample is then randomly generated by filling each unit cell either by In or by Ga corresponding to a potential energy of either 0 (In) or $\Delta E_c$ (Ga) for electrons and either 0 (In) or $\Delta E_v$ (Ga) for holes. For the quantum dots a complete 3D diagonalization was undertaken enclosing the pyramid in a box with a square basis with a side of 14.9~nm and a height of 7.9~nm. The box is then subdivided into smaller boxes of atomic size where the potential is supposed to be constant. The wave functions are then developed on the sinus solutions leading to a Hilbert space of dimension 75000.

For the QDs, two situations have been considered : either uncorrelated sites where the probability that a given cell is occupied by In is $x$ ( = 17\%) irrespective of the occupancy of the other sites or site correlations which favor the In cluster formation \cite{Narakawa, Donnell} by allowing a probability $p > x$ for a cell to be occupied by an In if its first neighborings are already occupied by at least one In. The only samples retained in the analysis are those such that the In fraction in the whole QD lies in the interval $[0.17, 0.18]$.

The first few eigenstates of those 2D (40 states) and 3D (6 states) problems are evaluated by means of exact diagonalizations using the L\'anczos algorithm.  Once the eigenstates are known one can calculate various averages like the mean position, the mean square deviation to it, etc..  

While the properties of a single sample have virtually no chance to be compared to any experiment, the average properties, obtained by taking the arithmetic mean of the outcomes of a given physical quantity in a series of samples, have more physical substance since they can (in principle) be compared to the results obtained on ensembles of quantum dots or quantum wells.  This is for these average properties that one can discuss the effectiveness or failure of the VCA.  One important quantity is $D_e$, the average electron density of states which is~:
\begin{eqnarray}\label{AverageElectronDOS}
D_e(\epsilon)=\frac{1}{\pi N}\sum_{j=1}^{N}\sum_{i}\frac{\Gamma/2}{\left(\epsilon-\epsilon_{ei}\left(j\right)\right)^2 + \left(\Gamma/2\right)^2}
\end{eqnarray}
where N is the number of samples (which differ from one another by the locations of the In atoms) while $\epsilon_{ei}(j)$ is the $i^{th}$ eigenvalue of the $j^{th}$ sample and $\Gamma$ has been taken equal to 1~meV. Here we replace the usual delta peaks by Lorentzian distributions for clarity. Equally important (and measurable) is the average electron - hole optical density of states defined as~:
\begin{eqnarray}\label{AverageElectronHoleOpticalDOS}
D_{eh}(\epsilon)=\frac{1}{\pi N}\sum_{j=1}^{N}\sum_{k,l}\frac{\left(\Gamma/2\right)\left|\langle\Psi_{ek}|\Psi_{hl}\rangle\right|^2}{\left(\epsilon-\epsilon_{ek}(j)-\epsilon_{hl}(j)\right)^2 + \left(\Gamma/2\right)^2}
\end{eqnarray}
In the finite QW structure the VCA results in peaks related to the confinement box which fulfill :
\begin{eqnarray}\label{EnergyLevel}
\epsilon_{np}=V_0+\epsilon_{1z}+\frac{\hbar^2 \pi^2}{2m_{//}L^2}\left(n^2+p^2\right)&;&n,p=1,2,...
\end{eqnarray}
where $V_0 = x \Delta E_c$ is the VCA average potential, $\epsilon_{1z}$ is the lowest confining energy in the Z direction and L = 36~nm.

\begin{figure}[!htbp]
\begin{center}\includegraphics[width=1.0\columnwidth,
  keepaspectratio]{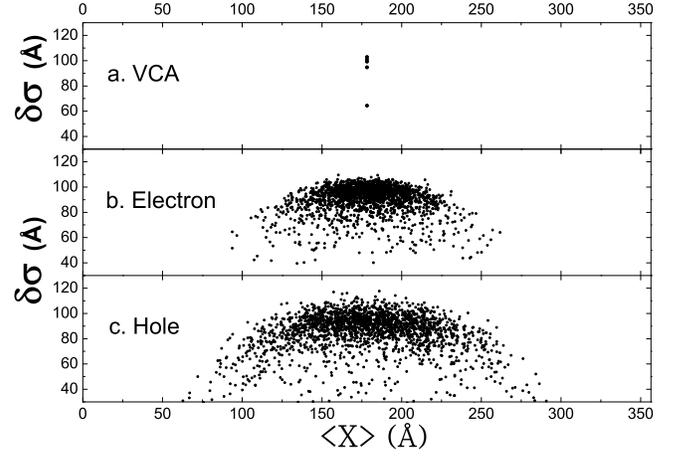}\end{center}
\caption{Calculated mean square deviations $\delta\sigma$ plotted versus mean carrier position $\langle{\rm X}\rangle$ for the VCA (a), electrons (b) and holes (c) in a finite ${\rm Ga}_{0.83}{\rm In}_{0.17}{\rm N}/{\rm GaN}$ quantum well.  Random alloys and 50~samples.}\label{MeanPosition}
\end{figure}

\begin{figure}[!htbp]
\begin{center}\includegraphics[width=1.0\columnwidth,
  keepaspectratio]{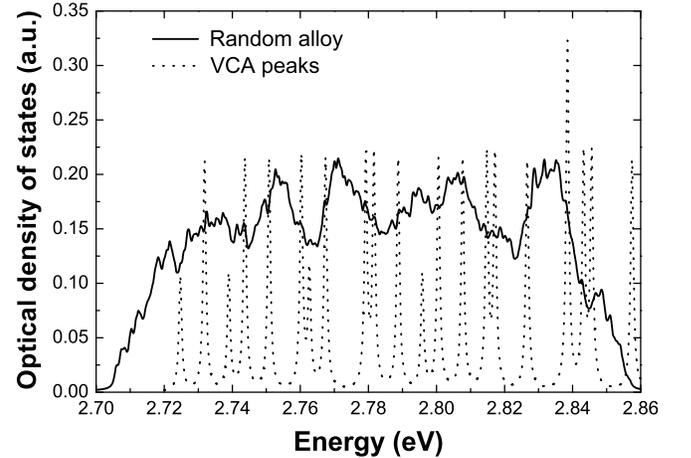}\end{center}
\caption{Optical density of states averaged over N = 50 ${\rm Ga}_{0.83}{\rm In}_{0.17}{\rm N}/{\rm GaN}$ quantum wells.  Random alloys.  Dashed lines : VCA peaks.}\label{OpticalDensity}
\end{figure}

Fig.(\ref{MeanPosition}) shows for the QW structure the mean square deviation $\delta \sigma$ plotted versus the mean location $\langle{\rm X}\rangle$ for the electron (upper panel) and the hole  (lower panel) along the X axis (the results are equivalent for the Y axis) for 50 samples, together with the VCA results. The In distribution is at random (${\rm x} = 0.17$).  It is obvious that $\langle{\rm X}\rangle$ and $\delta \sigma$ for holes differ strongly from the VCA predictions while the electrons with their lighter mass are less readily confined by In rich fluctuations and thus display $\delta\sigma(\langle{\rm X}\rangle)$ which are closer from the VCA results than the holes.  Fig.(\ref{OpticalDensity}) shows the electron - hole optical density of states $D_{eh}(\epsilon)$ versus the energy $\epsilon$ averaged over $N=50$  samples.  $D_{eh}$ is proportional to the light absorption coefficient if we neglect electron-hole interaction, which is justified here due to the strong e-h separation by the electric field (770 meV compared to the Coulomb interaction energy of the order of 44 meV).  The VCA results in a rounded (by $\Gamma$) staircase - like $D_{eh}$ for an infinite QW while the large box produces VCA peaks whose position follows eq. (\ref{EnergyLevel}).  We see that the random alloy results in a broadening of these peaks which is quite important.  Note for instance the significant ($D_{eh} > 0.05$) band tail which develops down to  $\approx 15$~meV below the nominal edge of the ground peak.  This bandtail corresponds to increasingly localized states around In clusters, in particular for the holes. The high energy decrease of the optical density of states is unphysical. It reflects the high energy cut-off of the electron  and hole eigenvalues. Notice that the photoluminescence line is found experimentally at $\simeq$2.6~eV in ${\rm Ga}_{0.83}{\rm In}_{0.17}{\rm N}/{\rm GaN}$ \cite{Morel1} with which our results are coherent.

\begin{figure}[!htbp]
\begin{center}\includegraphics[width=1.0\columnwidth,
  keepaspectratio]{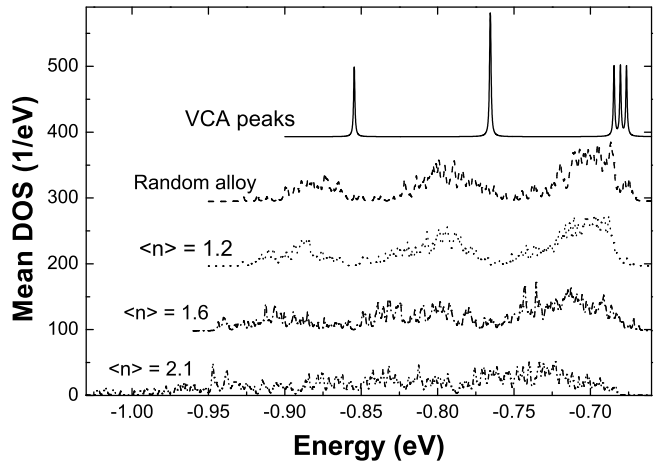}\end{center}
\caption{Electron density of states averaged over N = 100 ${\rm Ga}_{0.83}{\rm In}_{0.17}{\rm N}/{\rm GaN}$  pyramids. $\langle n\rangle$ is the mean number of first In neighbors for an In. From top to bottom: VCA results (divided by 6),  random alloy $\langle n\rangle=1.0$, and increasingly segregated alloys $\langle n\rangle=1.2$ ; $\langle n\rangle=1.6$ ;  $\langle n\rangle=2.1$. The energy zero is taken at the barrier monolayer next to the bottom of the wetting layer.}\label{DOSElectronN}
\end{figure}

\begin{figure}[!htbp]
\begin{center}\includegraphics[width=1.0\columnwidth,
  keepaspectratio]{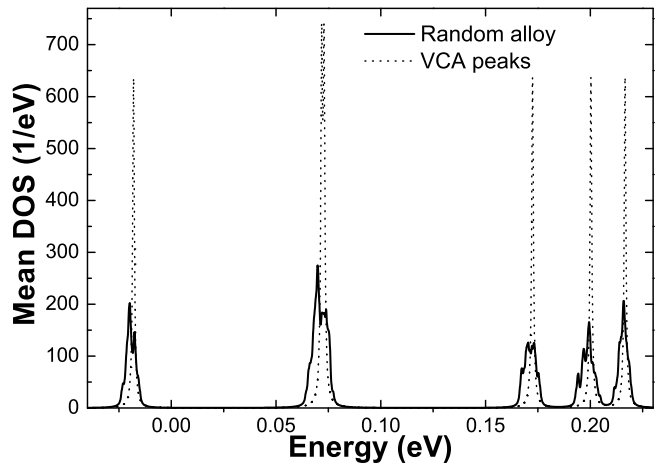}\end{center}
\caption{Electron density of states averaged over N = 50 ${\rm Ga}_{0.5}{\rm In}_{0.5}{\rm As}/{\rm GaAs}$ pyramids.  Solid lines :  random alloy. Dashed lines:VCA results.}\label{DOSElectronInAs}
\end{figure}

\vspace{3cm}

Fig.(\ref{DOSElectronN}) shows the calculated electron density of states for ${\rm Ga}_{0.83}{\rm In}_{0.17}{\rm N}/{\rm GaN}$ pyramid (100 samples); the optical density of states displays similar results although the hole states can be unbound in the pyramid. The VCA results shows a twofold degenerate excited state located some 90 meV above the ground state.  It lies some 80 meV below a narrow triplet.  The random alloy case shows both a redshift and a considerable broadening.  Note in particular the tendency towards the closing of the gaps between the remnants of the VCA peaks.  This effect is more pronounced when In segregation is introduced in the calculations. In addition, increasing the segregation produces densities of states which bear less and less resemblance with the VCA result. In contrast (fig.(\ref{DOSElectronInAs})) the ${\rm Ga}_{0.5}{\rm In}_{0.5}{\rm As}/{\rm GaAs}$ pyramids (50 samples) display densities of states which are very close to the VCA results.  Even though the 50\% alloy should display the maximum of disorder, we see that the calculated DOS features are very well separated from each others, and much narrower than the ones found in the 17\% nitride alloy. Note in particular that the bandtail extends only a few meV's below the VCA peaks.  Thus, if one knows the shape of a ${\rm Ga}_{0.5}{\rm In}_{0.5}{\rm As}$ pyramid, fig.(\ref{DOSElectronInAs}) shows that it does make sense to attempt to identify the peaks, to fit their energy difference etc.  On the other hand, the huge deviations from the VCA results in the 17\% nitride dot demonstrate that any peak attribution in this system is rather elusive.

\begin{acknowledgments}
We would like to thank Pr. Runge and Pr. Lef\`ebvre for fruitful discussions and insights.  The LPA-ENS is « Laboratoire associ\'e au CNRS et aux Universit\'es Paris 6 et Paris 7 ». 
\end{acknowledgments}

\end{document}